\documentclass{icrc29}
\usepackage{graphicx,amssymb,amsmath,times}
\begin{document}
\newenvironment{my_enumerate}
{\begin{enumerate}
    \setlength{\topsep}{1pt}
    \setlength{\partopsep}{0pt}
    \setlength{\itemsep}{0pt}
    \setlength{\parskip}{0pt}
    \setlength{\parsep}{0pt}
    }
{\end{enumerate}}
%
%
\title[Search for Correlated High Energy Cosmic Ray Events with CHICOS]
{Search for Correlated High Energy Cosmic Ray Events with CHICOS}
\author[R.~D.~McKeown et al.]
{R.~D.~McKeown$^a$,  B.E. Carlson$^a$, E. Brobeck$^a$, C.J.
Jillings$^a$, M.B. Larson$^a$, T.W. Lynn$^a$,
\newauthor
J.E. Hill$^b$, B.J. Falkowski$^c$, R. Seki$^c$, J. Sepikas$^d$, G.B.
Yodh$^e$.
\\
        (a) W.K. Kellogg Radiation Laboratory, California Institute of Technology,
        Pasadena, CA  91125, USA.
          \\
        (b) Department of Physics, California State University at Dominguez Hills,
        Carson, CA  90747, USA.
        \\
        (c) Department of Physics and Astronomy, California State
        University at Northridge, Northridge, CA  91330, USA.
        \\
        (d) Department of Astronomy, Pasadena City College,
        Pasadena, CA 91106, USA.
        \\
        (e) Department of Physics and Astronomy, University of California at Irvine,
        Irvine, CA  92697-4575, USA.
        }
\presenter{Presenter: R. D. McKeown (bmck@krl.caltech.edu),
usa-mckeown-R-abs2-he13-oral}

\maketitle

\begin{abstract}
We present the results of a search for time correlations in high
energy cosmic ray data (primary $E > 10^{14}$~eV) collected by the
California HIgh school Cosmic ray ObServatory (CHICOS) array. Data
from 69 detector sites spread over an area of 400 km$^2$ were
studied for evidence of isolated events separated by more than 1 km
with coincidence times ranging from 1 microseconds up to 1 second.
We report upper limits for the coincidence probability as a function
of coincidence time.
\end{abstract}

\section{Introduction}
Correlations between spatially separated cosmic ray events would
indicate that the particles have some common history. Extensive air
showers are well-known to produce correlated signals on the earth's
surface over distances of several kilometers. However, it is
possible that isolated primary cosmic ray events separated by $>
1$~km could arrive in time coincidence. Such correlated cosmic ray
events could result, for example, from the photodisintegration of
heavy nuclei (\emph{i.e.}, iron) by solar photons
\cite{photodisintegration}. In a  previous study
\cite{carrelmartin}, there was some episodic evidence for time
correlations up to $10^{-3}$~s in events separated by $\sim 100$~km.
More recently, LAAS \cite{LAAS} searched for correlated events at
large distances $\sim 500$~km and found a few candidate events.
However, these events were consistent with interpretation as
accidental coincidences between uncorrelated events. We have studied
data obtained with the CHICOS array during January 2003 through June
2005 and searched for evidence of correlated air shower events
separated by $> 1$~km with energy threshold $10^{14}$ eV.

\section{CHICOS}

The California HIgh school Cosmic ray ObServatory (CHICOS) observes
cosmic ray induced air showers with an array of detector sites
located on school roofs in the Los Angeles area (Lat.~$34.1^\circ$,
Long.~$-118.1^\circ$, average 250~m above sea level). The sites are
separated by distances of typically $2-3$ kilometers, with the
overall array covering an area of $\sim 400$~km$^{2}$. During this
study, the number of operational sites increased from 31 to 66. Each
detector site contains two plastic scintillator detectors, separated
by $\sim 3$ meters. The detector typically has $\sim 1$~m$^2$ area
and $5-10$~cm thickness. Data are stored on local hard disk and
automatically transferred to Caltech via internet every night by the
computer located at each site.

``Trigger'' events are defined as those where both detectors at a
site record signals greater than 2 single vertical particles within
a 100~ns time window. Most trigger events are isolated single events
({\it i.e.}, no nearby sites are hit ) which are generated by low
energy showers with a threshold of about $10^{14}$~eV. The rate of
these triggers ( about 1000 per day per site) is comparable to
expectations based on the previously measured flux \cite{nagano} and
computer simulations of air showers with the AIRES code
\cite{AIRES}.

Large showers that generate a trigger at one site with coincident
single hits at several neighboring sites must have extremely high
energies of $> 10^{18}$~eV.  We do observe such large air shower
events with the CHICOS array, presently at a rate of about one per
month. However, even the largest air showers with primary energy $E
\sim 10^{20}$~eV would not generate 2 triggers in sites separated by
more than 1~km. In this work we search for double trigger events
where two sites, separated by more than 1~km, both record trigger
events within a certain coincidence time. Such a signal could
indicate the existence of isolated correlated cosmic ray events
separated in distance by up to 60~km (the largest distance between
two CHICOS sites).

\section{Correlation Analysis}
The data sample for this analysis is the time-stamped sequence of
trigger data from the operational sites, which forms a complete
record of all the events detected by the array with sufficient
energy to trigger a single site. In order to examine these data for
time correlations, a randomized data set was constructed  directly
from the real data, as in \cite{carrelmartin}, by offsetting the
sequence of triggers at each site by some integer number of seconds
relative to the other sites. Since a shift of several seconds is
small compared to the time for drift in the average trigger rate,
the randomized data should reproduce all aspects of the real data
associated with accidental coincidences. Deviations of correlations
observed in the real data relative to the randomized data could be
indications of significant correlations (\emph{i.e.}, not
accidental) in the data.

\begin{figure}[b!]
\includegraphics[clip=true,scale=.32,angle=-90]{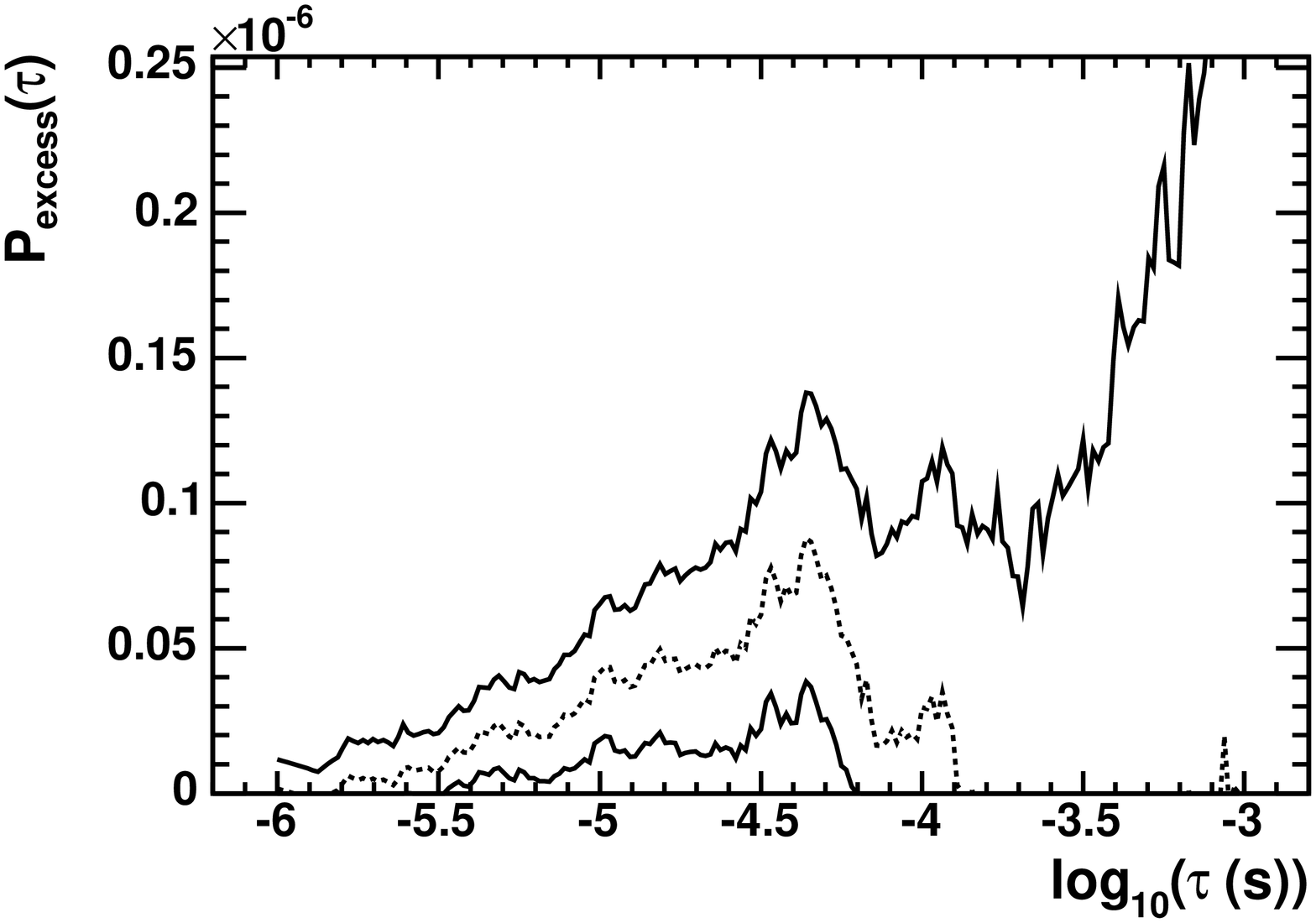}
\includegraphics[clip=true,scale=.32,angle=-90]{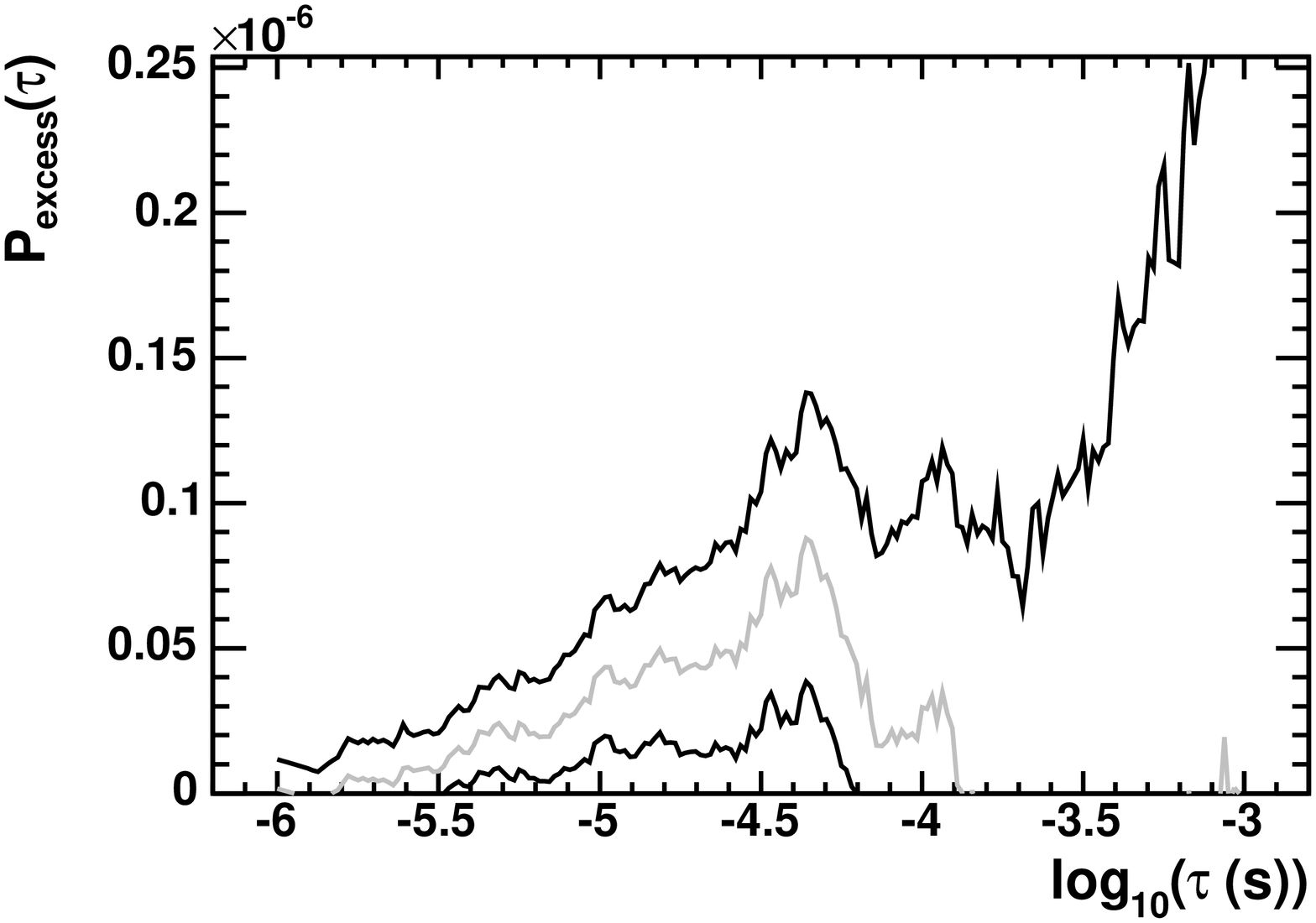}
\caption{Probability of an excess trigger coincidence per site
plotted vs. cumulative coincidence time scale. The upper plot shows
the results for short time scales $\tau< 10^{-3}$~s and the lower
plot shows the longer time scales $10^{-3}<\tau< 1$~s. The lines
drawn are the upper end (solid), center (dotted) and lower end
(solid) of a 90\% Feldman-Cousins confidence interval. The upper
curve is interpreted as an upper limit for the probability per site
of real coincidences for the corresponding time scale.}
\label{FCCIs}
\end{figure}

Given a coincidence time interval $\Delta t$, we define $N_{\rm
excess}$ as the number of excess event pairs (relative to the
randomized data sample) with that time difference, which may be
positive or negative. If both members of a successive pair are from
the same site, that pair is not counted (to eliminate instrumental
effects such as PMT afterpulsing).

In order to search for correlations on any time scale less than 1
second, we
compute the probability of an excess coincidence per site
for the cumulative time interval $\{ 0 , \tau \}$ according to
\begin{equation}
P_{\rm excess} (\tau) = \frac {N_{\rm excess} (\Delta t< \tau ) }
{N_{\rm trig}\, (\langle N_{\rm sites} \rangle - 1)}
\end{equation}
in which $\langle N_{\rm sites} \rangle$ is the average number of
operational sites and $N_{\rm trig}$ is the total number of trigger
pairs. We compute a 90\% confidence interval for $P_{\rm
excess}(\tau)$ using the method in \cite{feldmancousins}, and
interpret the upper limit as the 90\% confidence level upper limit
for the excess probability per site for the interval $\{ 0 , \tau
\}$. The results, for 17 months of data from January 2003 through
July 2004 are shown in Fig.~\ref{FCCIs} \cite{brant}.

\begin{figure}[b!]
\includegraphics[clip=true,scale=.32,angle=-90]{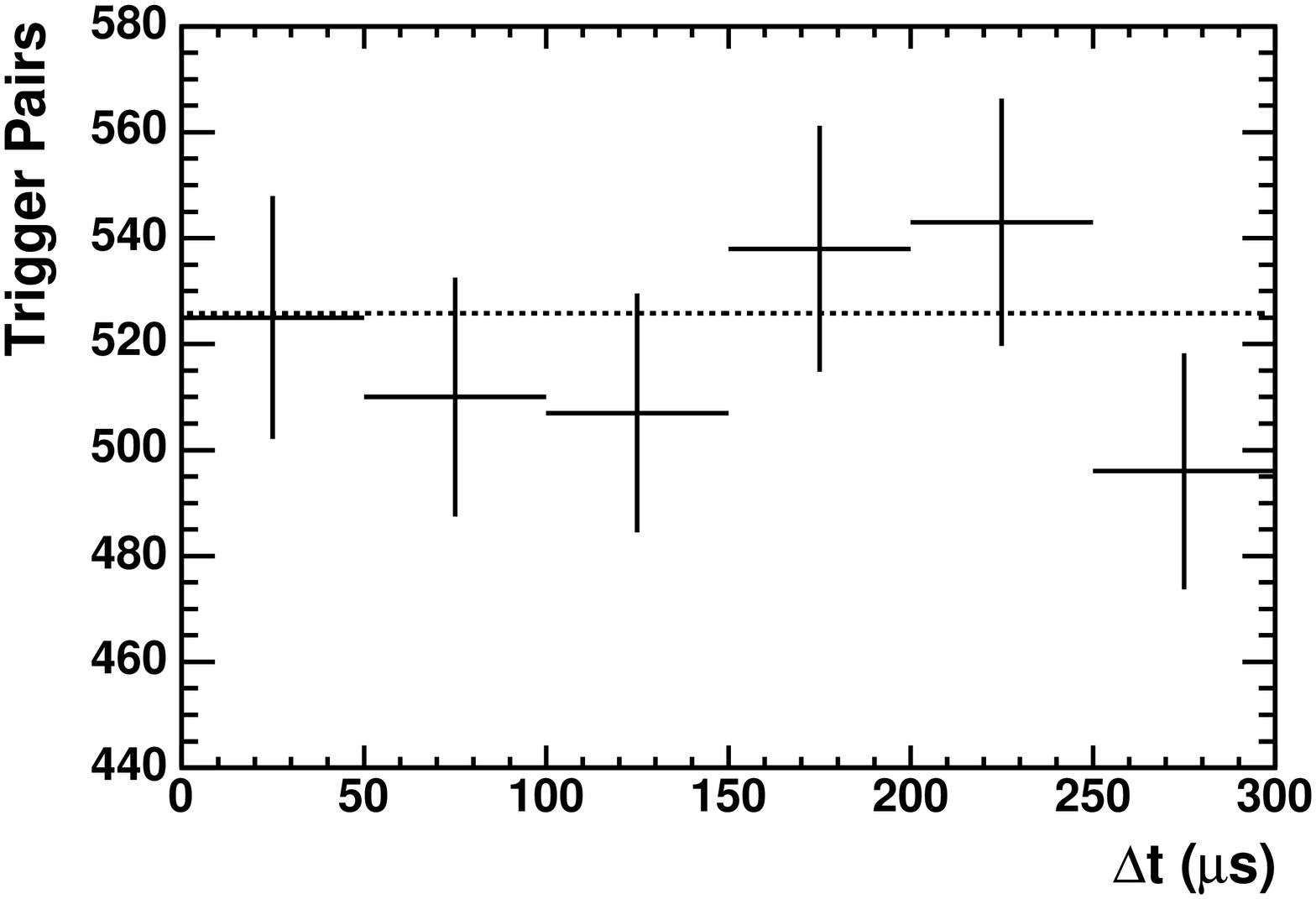}
\includegraphics[clip=true,scale=.32,angle=-90]{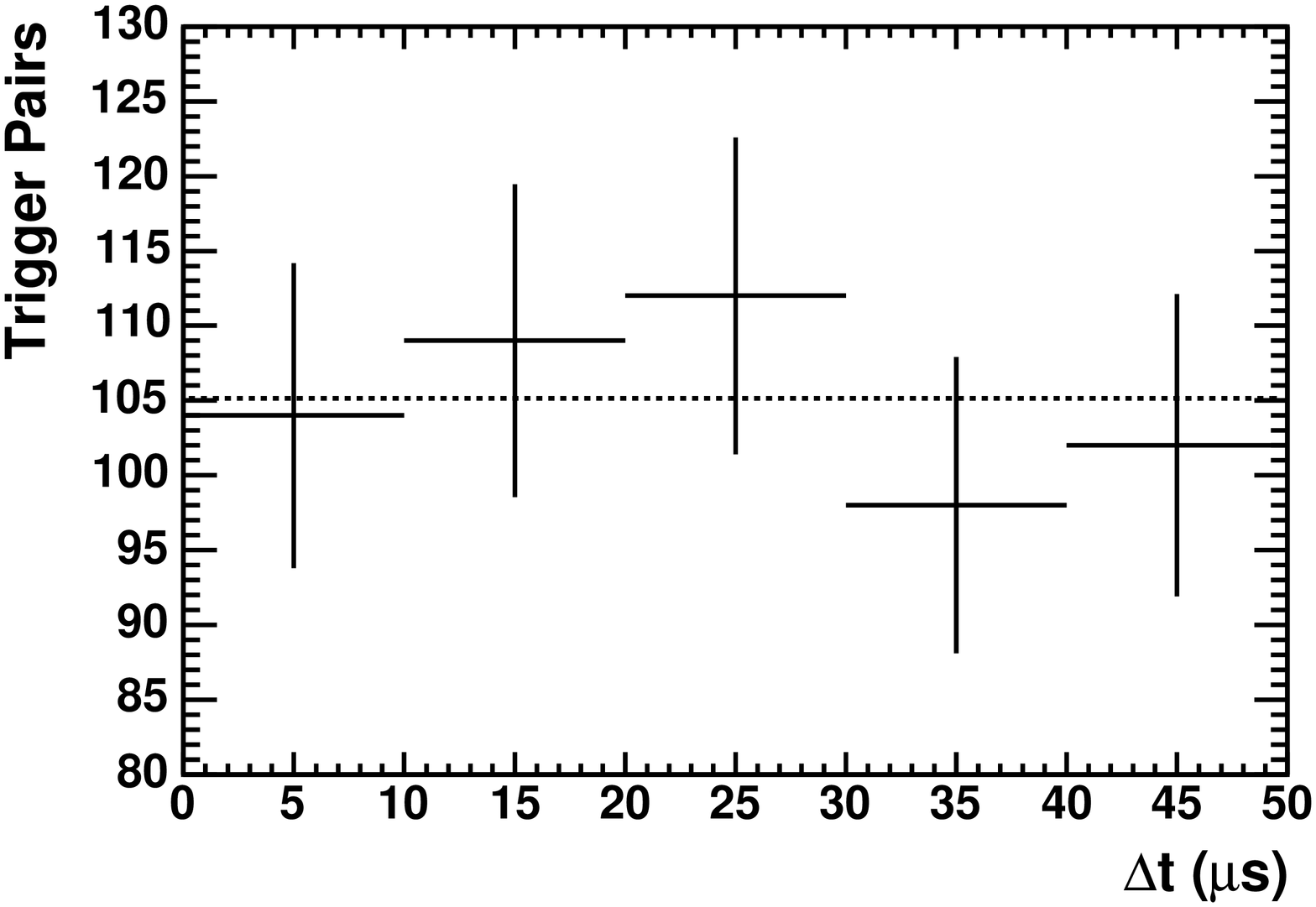}

\caption{Trigger pair time distribution for the new dataset (August
2004 through May 2005) with $\Delta t < 300 ~\mu$s and with $\Delta
t < 50 ~\mu$s. Plotted points with error bars are real data and the
horizontal dotted line indicates the expectation from random
coincidences.} \label{timediffs}
\end{figure}

As discussed in \cite{brant}, the data in  Fig.~\ref{FCCIs} indicate
an excess of events at time scales of order $\sim 10^{-4.5}$~s, or
$\sim 30 ~\mu$s. The excess heals itself at larger $\tau$ as we add
more data that show (apparently) no correlations. To study this
possible signal further, we have repeated this analysis with an
independent data set from August 2004 through May 2005. In
Fig.~\ref{timediffs} we show the results of the analysis of the new
data. It is clear from the new data that there is no evidence of the
apparent excess displayed in Fig.~\ref{FCCIs}. Therefore, the excess
in the earlier dataset is interpreted as a statistical fluctuation.

\section{Conclusions}

A search for time correlations in cosmic ray data collected by the
CHICOS project has been performed. The results are consistent with a
lack of any real correlation between isolated events. Earlier
observations that indicated an excess of events at shorter times are
interpreted as a statistical fluctuation. The excess coincidence
probabilities displayed in Fig.~\ref{FCCIs} can be properly
interpreted as 90\% CL upper limits for the incidence of correlated
pairs of cosmic rays above $10^{14}$~eV separated by distances
between 1 and 60~km.

\section{Acknowledgements}

We are grateful for the generous support of Caltech and the Weingart
Foundation in initiating the CHICOS project. The donation of the
detectors by the CYGNUS collaboration is gratefully acknowledged.
Support from the NSF (grants PHY-0244899 and PHY-0102502) and the
donation of computers for the project by IBM Corporation are also
acknowledged. The volunteer efforts of many high school and middle
school teachers
\footnote{http://www.chicos.caltech.edu/collaboration/collaboration-list.html}
have been essential in the deployment and operation of the CHICOS
array, and we are delighted to acknowledge their participation.

\end{document}